\documentclass[useAMS,usenatbib]{mn2e}
\usepackage{graphicx}
\title[The binary nucleus of IC~4776]{The post-common-envelope binary nucleus of the planetary nebula IC~4776: Neither an anomalously long orbital period nor a Wolf-Rayet binary\thanks{Based on observations made with the Southern African Large Telescope (SALT) under programme 2017-1-MLT-010.}}
\author[Miszalski et al.]{B. Miszalski,$^{1,2}$\thanks{E-mail: brent@saao.ac.za} R. Manick,$^{3,1}$ H. Van Winckel$^{3}$ and J. Miko{\l}ajewska$^{4}$\\
$^{1}$South African Astronomical Observatory, PO Box 9, Observatory, 7935, South Africa\\
$^{2}$Southern African Large Telescope Foundation, PO Box 9, Observatory, 7935, South Africa\\
$^{3}$Institute of Astronomy, KU Leuven, Celestijnenlaan 200D, B-3001 Leuven, Belgium\\
$^{4}$Nicolaus Copernicus Astronomical Center, Polish Academy of Sciences, Bartycka 18, PL-00716 Warsaw, Poland\\
}
\begin{document}

\date{Accepted . Received ; in original form }

\maketitle
\begin{abstract}
   The orbital period distribution of close binary stars consisting of a white dwarf and a main-sequence star (WDMS) is a powerful observational constraint on population synthesis models of the poorly understood common-envelope (CE) interaction. Models have often struggled to reproduce the small number of post-CE WDMS binaries with anomalously long orbital periods greater than $\sim$4 d, though recent studies suggest that in longer period systems recombination energy may help contribute to the efficient ejection of the CE. Planetary nebulae (PNe) are an emerging source of rare long period post-CE binaries which can act as powerful complementary constraints on population synthesis models to more traditional post-CE binary populations. A tentative 9.0 d orbital period was recently proposed for the central star of the PN IC~4776, potentially one of the longest periods observed in post-CE WDMS binaries. Here we present SALT HRS observations of IC~4776 that rule out a 9.0 d orbital period, as well as the previously suggested Wolf-Rayet classification of the primary. The SALT HRS data establish a 3.11 d orbital period and rule out Of and Wolf-Rayet primary spectral types. Assuming a mass of 0.6 $M_\odot$ for the primary and an orbital inclination matching the nebula orientation, we find a companion mass of $0.22\pm0.03$ $M_\odot$, most likely corresponding to an M4V companion. The orbital period of IC~4776 is still consistent with findings of abundance discrepancy factor (ADF) studies of post-CE PNe, but any trends in the ADF distribution derived from the sample remain significantly biased by selection effects.
\end{abstract}


\section{Introduction}
\label{sec:intro}
Planetary nebulae with close binary central stars are the most immediate ($\sim$10$^4$ yr) aftermath of the poorly understood common-envelope (CE) interaction (Ivanova et al. 2013). Their orbital period distribution is a potentially powerful constraint on CE population synthesis models (Miszalski et al. 2009) that aim to reproduce the present-day population of post-CE binaries (e.g. Davis et al. 2010; Nie et al. 2012; Toonen \& Nelemans 2013; Zorotovic et al. 2014a). Most post-CE central stars have main-sequence (MS) companions, making them the most useful to compare against models due to their greater numbers, however their overall numbers are still relatively small compared to other post-CE populations. The most numerous and well-studied post-CE population used to constrain CE models are more evolved binaries consisting of a white dwarf (WD) and a MS companion (WDMS binaries, e.g. Nebot G\'omez-Mor\'an et al. 2011). Many of these may have evolved from post-CE central stars and we may therefore expect post-CE central stars to exhibit a similar orbital period distribution to the bias-corrected WDMS distribution (Nebot G\'omez-Mor\'an et al. 2011). Indeed, previous comparisons show generally good agreement between the observed orbital period distributions (Miszalski et al. 2009; Hillwig 2011).

A recurrent challenge faced by CE models has been to explain the apparent lack of post-CE binaries with orbital periods of several days to weeks (e.g. Davis et al. 2010; Zorotovic et al. 2010; Nie et al. 2012; Toonen \& Nelemans 2013). This does not seem to be the result of observational biases, but rather that a low ejection efficiency of the CE is required (Zorotovic et al. 2010; Toonen \& Nelemans 2013; Camacho et al. 2014). Indeed, Nebot G\'omez-Mor\'an et al. (2011) found no observational evidence for a significant population of binaries with orbital periods of days to weeks, with the bias-corrected orbital period distribution of WDMS binaries encompassing orbital periods of 1.9 h to 4.3 d. 

There are, however, some post-CE central stars that have orbital periods exceeding this range (Tab. \ref{tab:long}) and these could therefore be considered anomalous in comparison to WDMS binaries. There are also a few post-CE WDMS binaries with similarly long orbital periods without PNe, e.g. KOI-3278 ($P$=88.18 d, Kruse \& Agol 2014), IK~Peg ($P$=21.72 d, Vennes et al. 1998), and SDSS J222108.45$+$002927.7 and SDSS J121130.94$-$024954.4 ($P$=9.59 d and $P$=7.82 d, respectively; Rebassa-Mansergas et al. 2012). Whether there exists a large population of anomalously long orbital period post-CE binaries remains to be determined. Several studies have suggested that an additional energy source, such as recombination energy, is required to contribute towards efficient CE ejection in these binaries (Han et al. 1995; Rebassa-Mansergas et al. 2012; Zorotovic et al. 2014a, 2014b; see also Nandez \& Ivanova 2016 and Iaconi et al. 2017). 

\begin{table}
   \centering
   \caption{Binary central stars of PNe with orbital periods $P$ longer than 4.0 d. The third column indicates whether the binary experienced a CE interaction phase.}
   \label{tab:long}
   \begin{tabular}{lrll}
      \hline\hline
      Name & $P$ (d) & Post-CE? & Ref.\\
      \hline
      NGC~1514 & $3306\pm60$ & N & (1) \\
      LoTr~5 & $2717\pm63$ & N & (1,2) \\
      PN G052.7$+$50.7 & $1105\pm24$ & N & (2)\\
      NGC~1360 & $141.6\pm0.8$ & Y? &   (3) \\
      MyCn~18 & $18.15\pm0.04$ & Y & (4)\\
      NGC~2346 & $16.00\pm0.03$ & Y & (5,6) \\
      NGC~5189         & $4.05\pm0.10$ & Y  & (7,8)\\
      \hline\hline
   \end{tabular}
      \begin{flushleft}
         \footnotesize{References: (1) Jones et al. (2017); (2) Van Winckel et al. (2014); (3) Miszalski et al. (2018a); (4) Miszalski et al. (2018b); (5) M\'endez \& Niemela (1981); (6) Brown et al. (2018); (7) Manick et al. (2015); (8) Miszalski et al. (2015).}

      \end{flushleft}
\end{table}

Zorotovic et al. (2014a) incorporated varying contributions of the recombination energy to CE ejection in simulations that predict orbital period distributions of post-CE binaries. The results of the simulations are particularly encouraging when it comes to explaining long orbital period post-CE binaries, in some cases producing orbital periods up to $\sim$1000 d. However, placing observational constraints on the Zorotovic et al. (2014a) predictions is currently hampered by the acknowledged bias in the WDMS sample towards M-dwarf companions and their associated short orbital periods (Nebot G\'omez-Mor\'an et al. 2011). Parsons et al. (2016) plan to ameliorate this bias by expanding the WDMS sample to include earlier spectral type companions that are expected to have longer orbital periods. Rebassa-Mansergas et al. (2017) determined a preliminary close binary fraction of this sample to be $\sim$10 per cent.

Apart from the Parsons et al. (2016) sample, the central stars of PNe offer an excellent complementary population with which to probe the orbital period distribution of WDMS post-CE binaries and the associated selection effects. Indeed, some of the longest orbital periods of post-CE binaries are already found in PNe (Tab. \ref{tab:long}). We are further expanding and characterising the population of long period binary central stars via a systematic radial velocity (RV) monitoring survey with the Southern African Large Telescope (SALT, Buckley, Swart \& Meiring 2006; O'Donoghue et al. 2006) High Resolution Spectrograph (Miszalski et al. 2018a, 2018b).

In this context we are also investigating candidates with uncertain or suspected long orbital periods like MPA~1508$-$6455 (Miszalski et al. 2011) and IC~4776 (Sowicka et al. 2017). Miszalski et al. (2011) found a potential orbital period of 12.5 d from photometry of the central star of MPA~1508$-$6455, but SALT HRS measurements prove that its orbital period is considerably longer (Bonokwane 2018).\footnote{A detailed study of MPA~1508$-$6455 will be presented elsewhere.} Sowicka et al. (2017) recently identified peak-to-peak RV variability of 80 km s$^{-1}$ in the central star of IC~4776 (PN G002.0$-$13.4). Qualitative analysis of the ten RV measurements by Sowicka et al. (2017) favoured a tentative 9.0 d orbital period, which would constitute an anomalously long orbital period for WDMS binaries (Nebot G\'omez-Mor\'an et al. 2011) and place it amongst the longest in PNe (Tab. \ref{tab:long}). 

Despite its unproven binary nature and uncertain orbital period, suspected to lie between 1.0 and 20.0 d (Sowicka et al. 2017), the tentative 9.0 d period was used by Sowicka et al. (2017) and Wesson et al. (2018) to make inferences about the potential dependence of the abundance discrepancy factors (ADF) of post-CE PNe with the orbital period. The ADF is usually calculated as the ratio of the oxygen abundance determined from recombination lines to that determined from collisionally excited lines and any such trends are helpful to understand the mysterious origin of `extreme' ADFs where the ADF exceeds a value of 10 (see Corradi et al. 2015; Wesson et al. 2018), but the orbital period is particularly significant as shorter orbital periods may help facilitate nova-like activity that is the current favoured explanation for `extreme' ADFs (Wesson et al. 2018). 

The spectroscopic classification of the central star of IC~4776 also remains uncertain. Aller \& Keyes (1985) and Feibelman et al. (1999) assigned Wolf-Rayet classifications (e.g. Crowther et al. 1998; Acker \& Neiner 2003), however these were not upheld by UV spectroscopy (Herald \& Bianchi 2004). Although Sowicka et al. (2017) discussed IC~4776 in the context of Wolf-Rayet binary central stars, there are only two unambiguous examples known, namely PN G222.8$-$04.2 and NGC~5189 (Hajduk et al. 2010; Manick et al. 2015; Miszalski et al. 2015). It is important to clarify the spectral type of IC~4776 as any new examples of Wolf-Rayet binary central stars would be potentially very helpful to clarify the unclear formation of Wolf-Rayet central stars in PNe (e.g. Todt et al. 2010; Miszalski et al. 2012; Todt et al. 2013).

To clarify all these issues surrounding IC~4776 we were motivated to include it in our ongoing SALT HRS program. In this paper we report on the results of the SALT HRS observations which are presented in Sect. \ref{sec:obs}. We analyse and discuss the measurements in Sect. \ref{sec:results} before concluding in Sect. \ref{sec:conclusion}. 

\section{SALT HRS OBSERVATIONS}
\label{sec:obs}
\begin{figure*}
   \begin{center}
      \includegraphics[scale=0.35,bb=0 0 642 457]{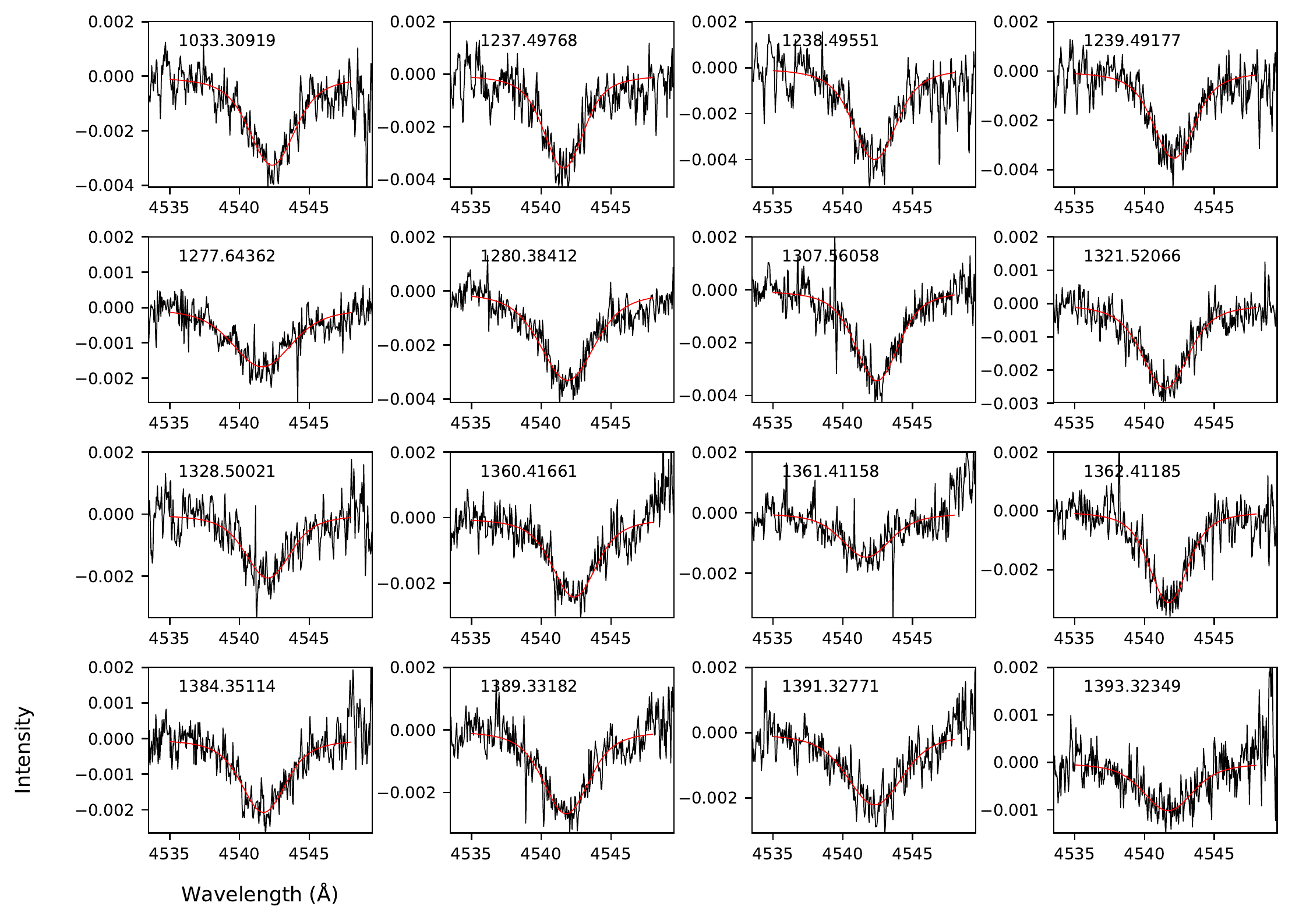}
      \includegraphics[scale=0.35,bb=0 0 629 457]{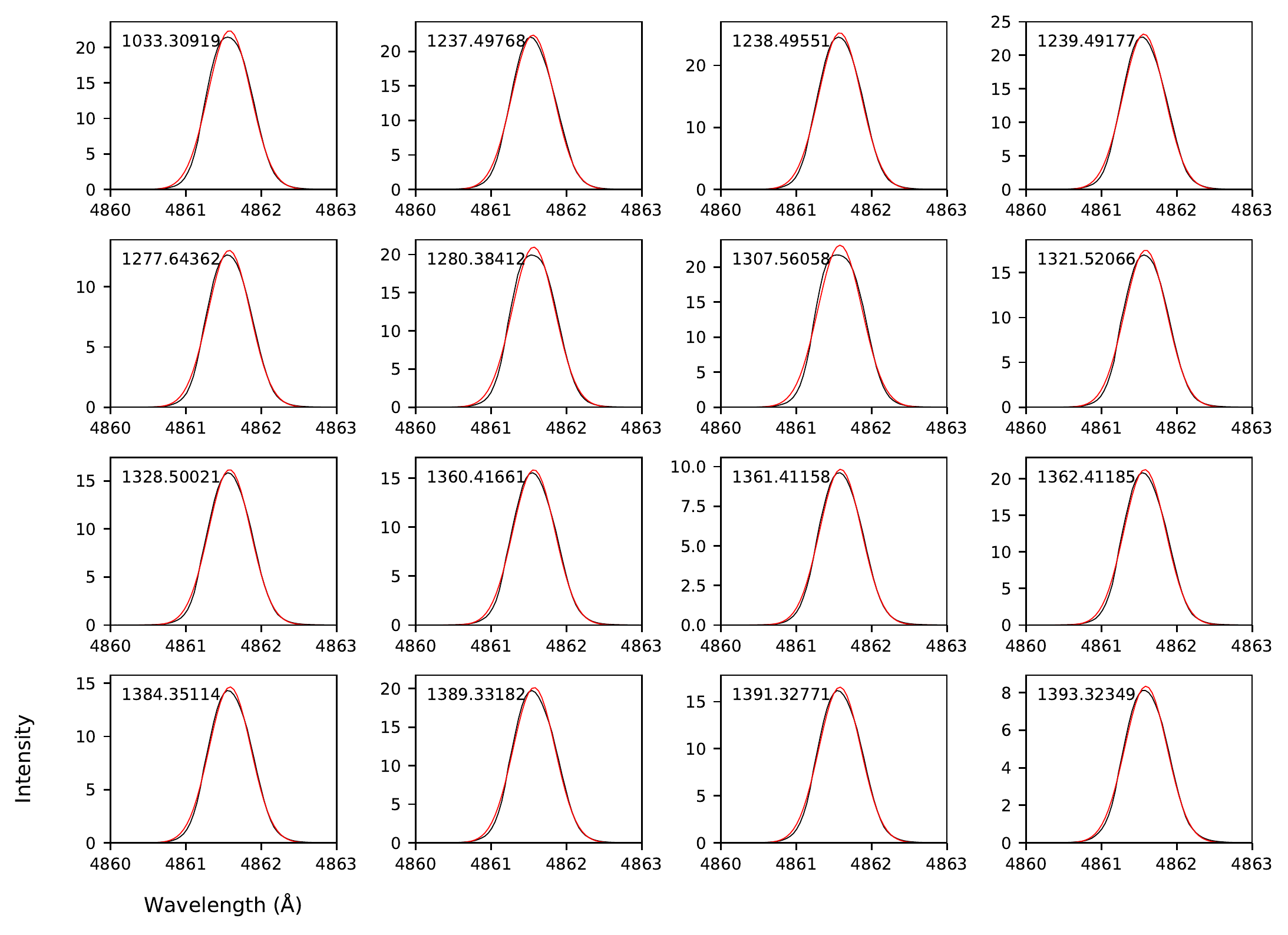}
   \end{center}
   \caption{\emph{(Top)} The observed stellar He~II $\lambda$4541.59 \AA\ profiles (black lines) and the Voigt function fits (red lines). \emph{(Bottom)} The observed nebular H$\beta$ $\lambda$4861.36 \AA\ profiles (black lines) and the Gaussian function fits (red lines). Each panel is labelled with the Julian day of each spectrum minus 2457000 days.}
   \label{fig:fits}
\end{figure*}

\begin{figure*}
   \begin{center}
      \includegraphics[scale=0.75,bb=0 0 603 256]{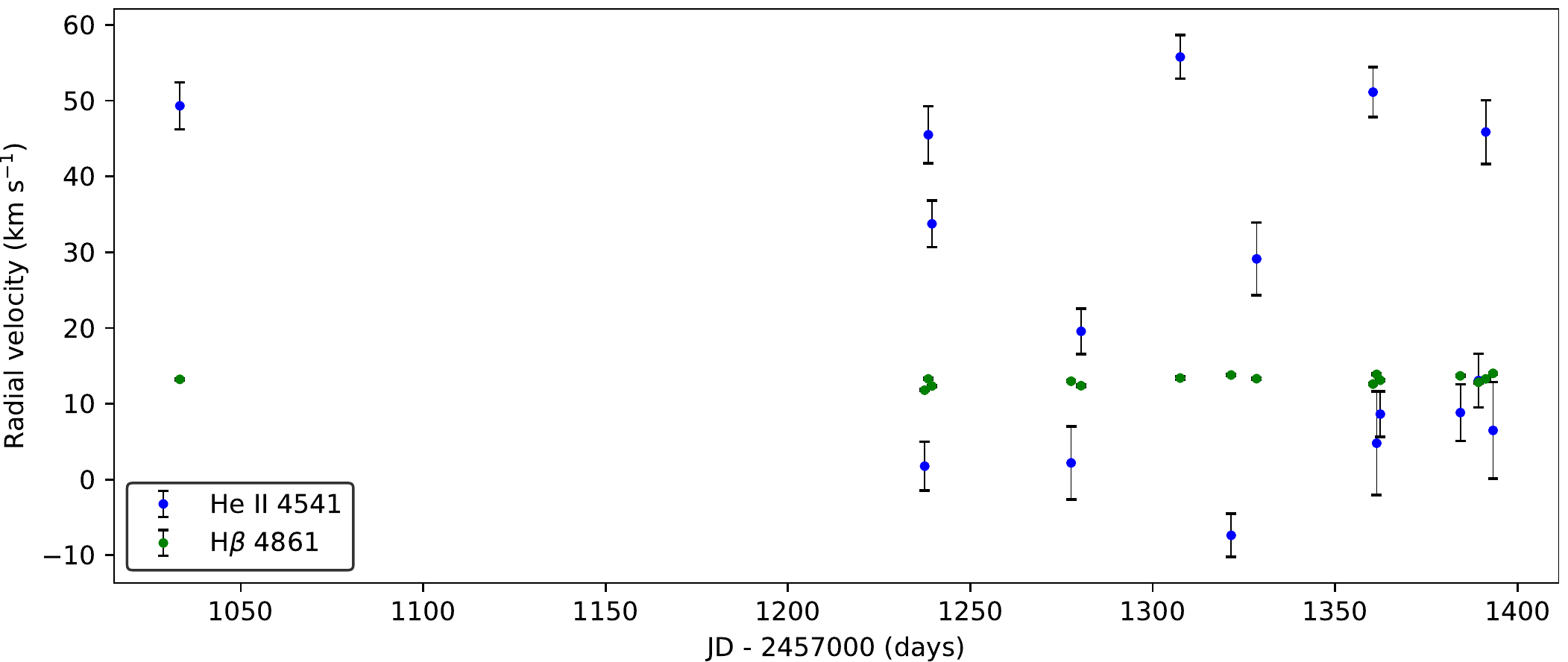}
   \end{center}
   \caption{The SALT HRS RV measurements of He~II $\lambda$4541 (stellar) and H$\beta$ $\lambda$4861 (nebular). The uncertainties in the nebular measurements are smaller than the symbol size.}
   \label{fig:rvjd}
\end{figure*}

We obtained 16 \'echelle spectra of IC~4776 with the High Resolution Spectrograph (HRS) on SALT (Bramall et al. 2010, 2012; Crause et al. 2014) under programme 2017-1-MLT-010 (PI: Miszalski). Table \ref{tab:log} presents a log of the observations taken with the medium resolution mode. We use the blue arm data ($R=\lambda/\Delta\lambda=43000$) for RV measurements (for details see Miszalski et al. 2018a). After basic reductions (Crawford et al. 2010) the data were reduced with the \textsc{midas} pipeline of Kniazev et al. (2016) that is based on the \textsc{echelle} (Ballester 1992) and \textsc{feros} (Stahl et al. 1999) packages. We applied heliocentric corrections to the data using \textsc{velset} of the \textsc{rvsao} package (Kurtz \& Mink 1998). The heliocentric RV measurements in Table \ref{tab:log} were obtained by fitting single Voigt and Gaussian functions to stellar He~II $\lambda$4541.59  and nebular H$\beta$ $\lambda$4861.36 features, respectively, using the \textsc{lmfit} package (Newville et al. 2016). Figure \ref{fig:fits} shows the fits to the data and the RV measurements are depicted graphically in Fig. \ref{fig:rvjd}. We determine a mean heliocentric nebular RV of 13.1$\pm$0.6 km s$^{-1}$, where the uncertainty is the standard deviation of the individual measurements, in fair agreement with 16.3$\pm$0.6 km s$^{-1}$ reported by Durand et al. (1998). Sowicka et al. (2017) adopted the Durand et al. (1998) value as their systemic velocity and zero-point for their RV measurements. 

\begin{table}
   \centering
   \caption{Log of SALT HRS observations of IC4776. The Julian day represents the mid-point of each exposure and the heliocentric radial velocity measurements are made from stellar He~II $\lambda$4541 and nebular H$\beta$ 4861.}
   \label{tab:log}
   \begin{tabular}{lrrl}
      \hline\hline
      Julian day & Exposure & RV (He~II) & RV (H$\beta$)\\
                 & time (s) &  (km s$^{-1}$) & (km s$^{-1}$)\\
      \hline
 2458033.30919 & 2600 & 49.32$\pm$3.08  & 13.22$\pm$0.15\\
 2458237.49768 & 2650 & 1.76$\pm$3.22   & 11.79$\pm$0.12\\
 2458238.49551 & 2650 & 45.50$\pm$3.78  & 13.28$\pm$0.11\\
 2458239.49177 & 2650 & 33.75$\pm$3.07  & 12.34$\pm$0.11\\
 2458277.64362 & 2650 & 2.19$\pm$4.81   & 12.97$\pm$0.13\\
 2458280.38412 & 2650 & 19.56$\pm$2.99  & 12.38$\pm$0.17\\
 2458307.56058 & 2650 & 55.77$\pm$2.88  & 13.39$\pm$0.21\\
 2458321.52066 & 2900 & $-$7.37$\pm$2.86  & 13.79$\pm$0.13\\
 2458328.50021 & 2650 & 29.12$\pm$4.79  & 13.31$\pm$0.11\\
 2458360.41661 & 2650 & 51.13$\pm$3.30  & 12.60$\pm$0.12\\
 2458361.41158 & 3000 & 4.79$\pm$6.86   & 13.88$\pm$0.10\\
 2458362.41185 & 2450 & 8.63$\pm$2.99   & 13.10$\pm$0.13\\
 2458384.35114 & 2650 & 8.83$\pm$3.73   & 13.68$\pm$0.12\\
 2458389.33182 & 2650 & 13.06$\pm$3.53  & 12.83$\pm$0.13\\
 2458391.32771 & 2650 & 45.87$\pm$4.21  & 13.28$\pm$0.12\\
 2458393.32349 & 2650 & 6.48$\pm$6.38   & 14.01$\pm$0.12\\
      \hline
   \end{tabular}
\end{table}

\section{Analysis and Discussion}
\label{sec:results}

\subsection{Orbital parameters}
\label{sec:orbit}
\begin{figure}
   \begin{center}
      \includegraphics[scale=0.45,bb=0 0 576 432]{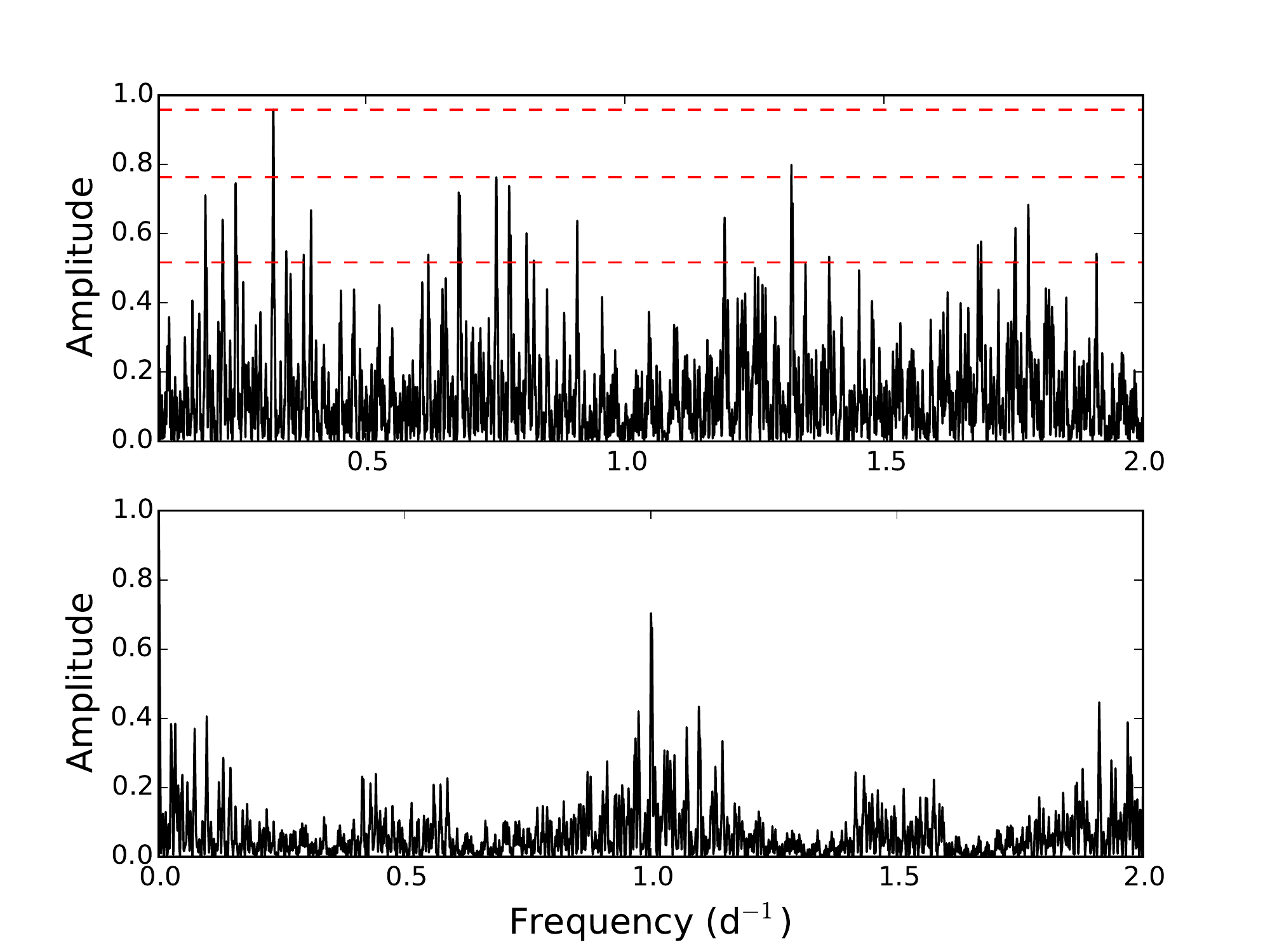}
      \includegraphics[scale=0.45,bb=0 0 576 432]{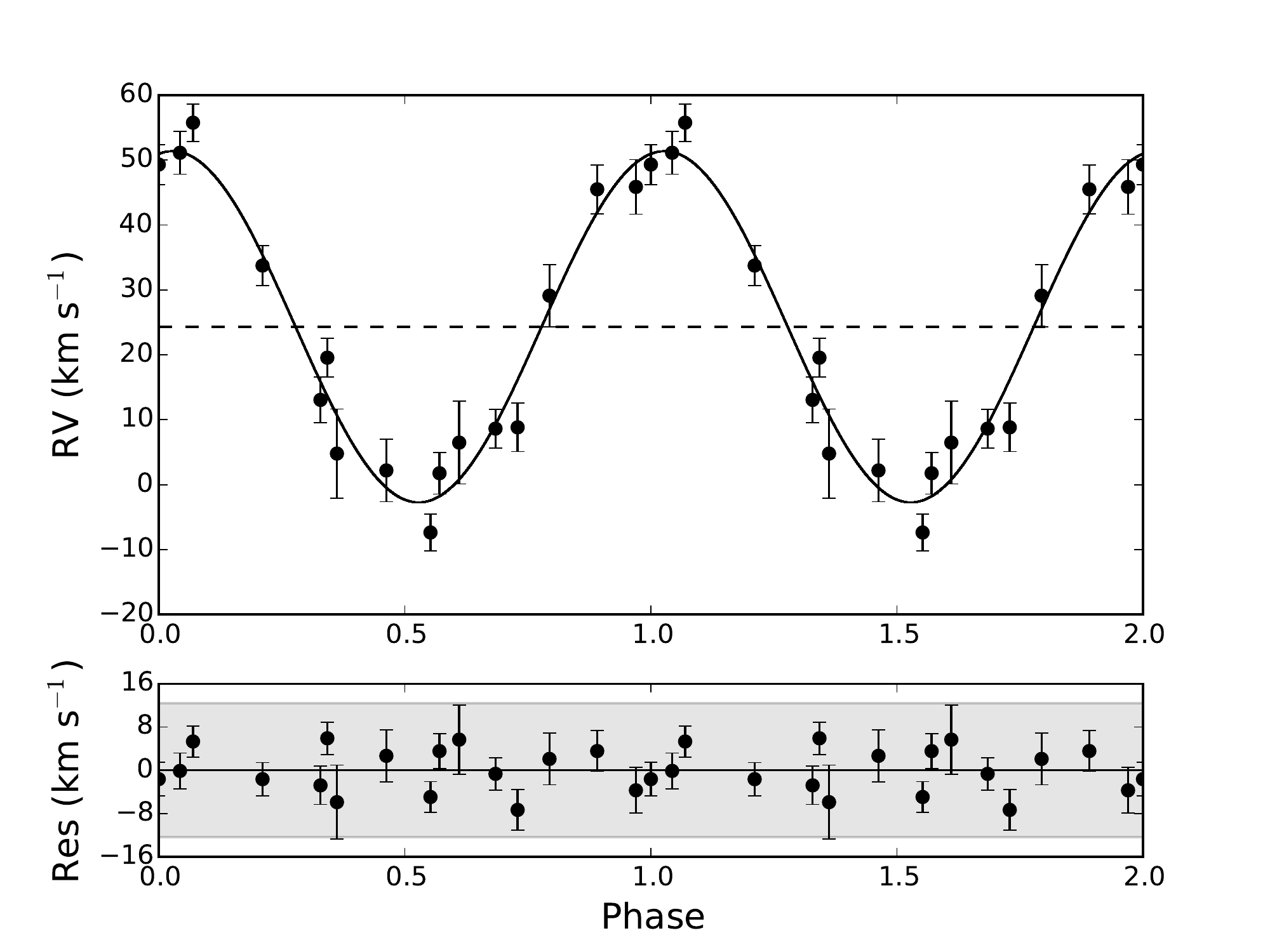}
   \end{center}
   \caption{\emph{(Top panel)} Lomb-scargle periodogram of the SALT HRS HeII $\lambda$4541 RV measurements (top half) and the window function (bottom half). The strongest peak at $f=0.321$ d$^{-1}$ corresponds to the orbital period and the dashed red lines correspond to 3, 4 and 5 sigma detection limits. \emph{(Bottom panel)} SALT HRS RV measurements phased with the orbital period. The solid line respresents the Keplerian orbit fit and the shaded region indicates the resdiuals are within 3$\sigma$ of the fit where $\sigma=4.1$ km s$^{-1}$.} 
   \label{fig:rv}
\end{figure}

We analysed the RV measurements in Tab. \ref{tab:log} using a Lomb-Scargle periodogram (Press et al. 1992). Figure \ref{fig:rv} shows the strongest peak in the periodogram to be at $f=0.321$ d$^{-1}$, corresponding to an orbital period of 3.11 d. The orbital period is significant at the 5$\sigma$ level and there is no significant peak corresponding to the 9.0 d period of Sowicka et al. (2017). We used the 3.11 d period as the basis for a Keplerian orbit model that was built using a least-squares minimisation method applied to the phase-folded data. Figure \ref{fig:rv} also shows the RV measurements phased with the orbital period, together with the Keplerian orbit fit and the residuals. 

Table \ref{tab:orbit} presents the orbital parameters of the fitted Keplerian orbit that were determined using Monte Carlo simulations in which the eccentricity was fixed to be zero (for details see Miszalski et al. 2018a). The systemic velocity $\gamma$ of $24.24\pm0.99$ km s$^{-1}$ differs from the nebular systemic velocity by a modest 11.1 km s$^{-1}$, which is not unexpected for He~II $\lambda$4541 in central stars such as IC~4776 that demonstrate a substantial wind (M\'endez et al. 1990). The terminal velocity of the stellar wind in IC~4776 is $v_\infty=2300\pm200$ km s$^{-1}$ (Herald \& Bianchi 2004) and comparable to some Of-type central stars (e.g. NGC~3242, Pauldrach et al. 2004), though the classification is not an Of-type (Sect. \ref{sec:class}).

\begin{table*}
   \centering
   \caption{Orbital parameters of the binary nucleus of IC~4776 derived from the best-fitting Keplerian orbit to HeII $\lambda$4541 measurements. The orbital inclination is assumed to match the nebula orientation (Sowicka et al. 2017).}
   \label{tab:orbit}
   \begin{tabular}{lc}
      \hline\hline
      Orbital period (d)        & 3.114$\pm$0.002\\
      Eccentricity $e$ (fixed)       & 0.00 \\
      Radial velocity semi-amplitude $K$ (km s$^{-1}$)  & 27.15$\pm$1.39 \\
      Systemic velocity $\gamma$ (km s$^{-1}$) & 24.24$\pm$0.99\\
      Epoch at radial velocity minimum $T0$ (d)            & 2458393.050$\pm$0.002\\
      Root-mean-square residuals of Keplerian fit (km s$^{-1}$) & 4.1 \\
      Separation of primary from centre of mass $a_1\sin i$ (au) & $0.00777\pm$0.00039 \\
      Mass function $f(M)$ (M$_\odot$) & $0.00648\pm$0.00098\\
      Inclination $i$ ($^\circ$) & $48\pm4$ \\
      \hline
   \end{tabular}
\end{table*}

The SALT HRS RV measurements quantitatively prove the post-CE binary nature of IC~4776, previously only hinted at based on the large RV variability measured by Sowicka et al. (2017). Unlike the tentative 9.0 d orbital period of Sowicka et al. (2017), the relatively short orbital period of 3.11 d is not considered anomalous in comparison with the bias-corrected orbital period distribution of WDMS binaries (Nebot G\'omez-Mor\'an et al. 2011). Thanks to the spatio-kinematic modelling of Sowicka et al. (2017), the nebula orientation of $48\pm4$ deg to the line of sight may be assumed to correspond to the orbital inclination as in other post-CE PNe (Hillwig et al. 2016), allowing us to estimate the companion mass as a function of the primary mass using the mass function in Tab. \ref{tab:orbit}. 

\begin{figure}
   \begin{center}
      \includegraphics[scale=0.55,bb=0 0 424 316]{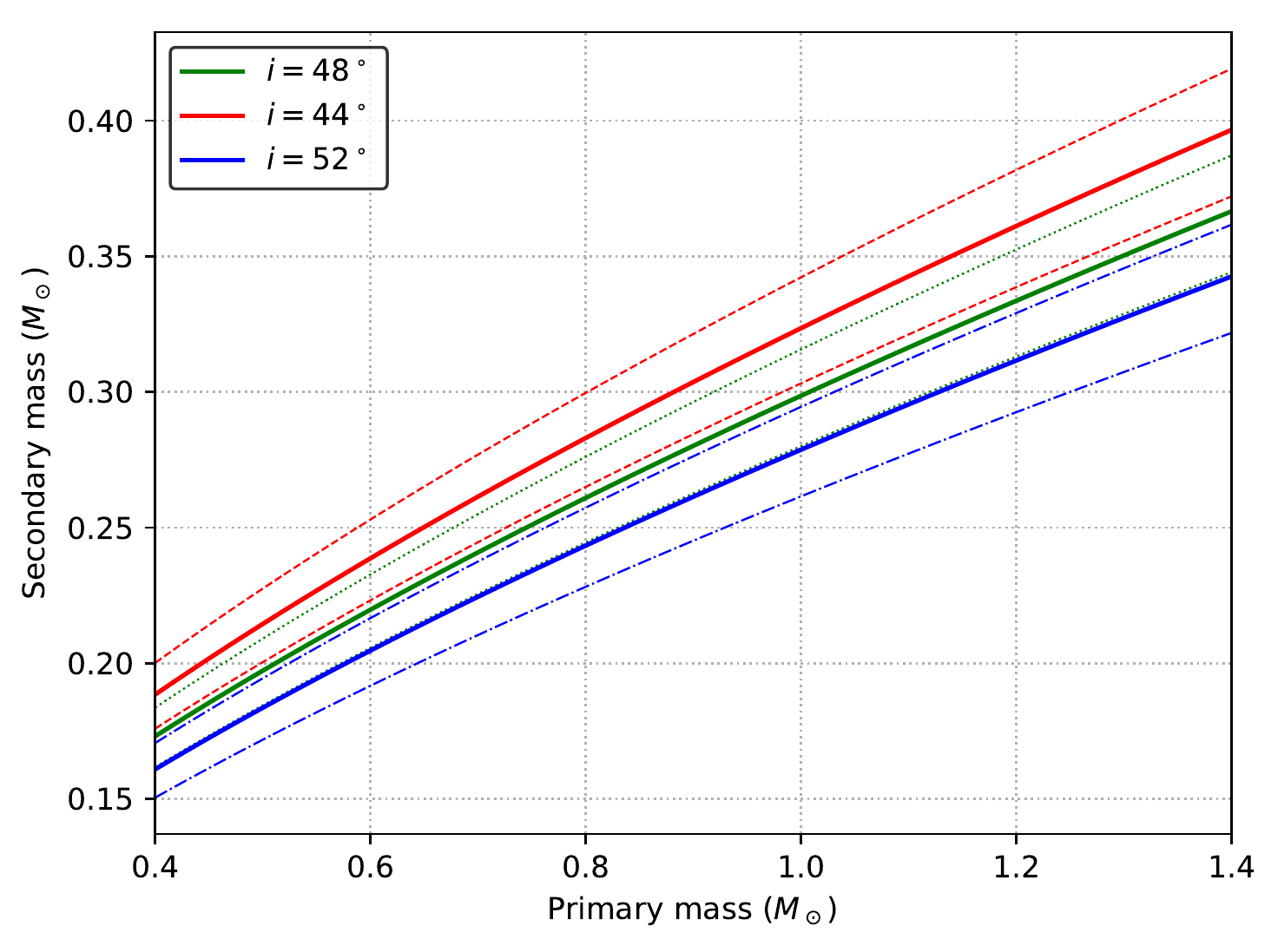}
   \end{center}
   \caption{Secondary masses permitted by the mass function for a range of primary masses. The solid lines correspond to the nebula inclination of $48\pm4$ deg (Sowicka et al. 2017) assumed to be the orbital inclination. The dashed, dotted and dash-dotted lines reflect the uncertainty in the mass function at each inclination.}
   \label{fig:masses}
\end{figure}

Figure \ref{fig:masses} shows the calculated range of permitted companion masses which immediately rules out a more massive WD companion than the primary. Herald \& Bianchi (2004) determined a primary mass of 0.57 $M_\odot$ from model atmosphere analysis of UV spectra. Since no uncertainty in the mass was given by Herald \& Bianchi (2004), for the purposes of discussion we assume the canonical WD value of 0.6 $M_\odot$, which according to Fig. \ref{fig:masses} would yield a companion mass of $0.22\pm0.03$ $M_\odot$. This mass most likely corresponds to an approximate spectral type of M4V with an uncertainty of one spectral type (Bressan et al. 2012; Chen et al. 2014; Rajpurohit et al. 2013). Assuming a primary mass of 0.6 $M_\odot$ and an M4V companion spectral type, the Roche lobe radius for the companion (Eggleton 1983) would be $0.67\pm0.04$ $R_\odot$, where the uncertainty reflects the uncertainty in the orbital inclination. The companion mass would correspond to a radius $\sim$0.24 $R_\odot$ (Bressan et al. 2012; Chen et al. 2014) and therefore the Roche lobe radius would only be filled if the companion radius were inflated by $\sim$2.6 times. The companion radii of several post-CE binaries may be found to be inflated to a similar degree (e.g. Af{\c s}ar \& Ibano{\v g}lu 2008), but photometric observations and modelling are required to further investigate this possibility in IC~4776. 

The nearest post-CE PNe in the ADF versus orbital period diagram to IC~4776 (ADF=1.75, Sowicka et al. 2017) are now NGC~2392 ($P$=1.9 d, Miszalski et al. 2019; ADF=1.65, Zhang et al. 2012) and NGC~5189 ($P$=4.05 d, Manick et al. 2015; Miszalski et al. 2015; ADF=1.6, Garc{\'{\i}}a-Rojas et al. 2013). The newly determined 3.11 d orbital period of IC~4776 does not dramatically alter the conclusions of Wesson et al. (2018), however we consider the sample size still too small to draw any correlation between ADF and orbital period. The sample is dominated by selection effects, one of them being a spatial bias due to the choice of slit location (Wesson et al. 2018). The dominant selection effect influencing the entire sample is that the majority of post-CE PNe with determined ADFs and orbital periods above one day were discovered from RV monitoring. 

\subsection{Spectroscopic classification}
\label{sec:class}
The classification of the central star of IC~4776 is very uncertain in the literature, with attempts to classify the central star based on detections of C~III $\lambda$2297 (Feibelman 1999) and C~III $\lambda$4650 (Aller \& Keyes 1985; Tylenda et al. 1993) emission lines. Herald \& Bianchi (2004) noted the absence of C~III emission lines in the UV spectrum of IC~4776, concluding this did not support the [WC6] Wolf-Rayet classification of Feibelman (1999). Sowicka et al. (2017) discussed IC~4776 in the context of Wolf-Rayet binary central stars even though the spectral type of the primary was very uncertain. The SALT HRS spectra allow us to revisit the spectral classification of IC~4776.

\begin{figure}
   \begin{center}
      \includegraphics[scale=0.35,bb=0 0 720 720]{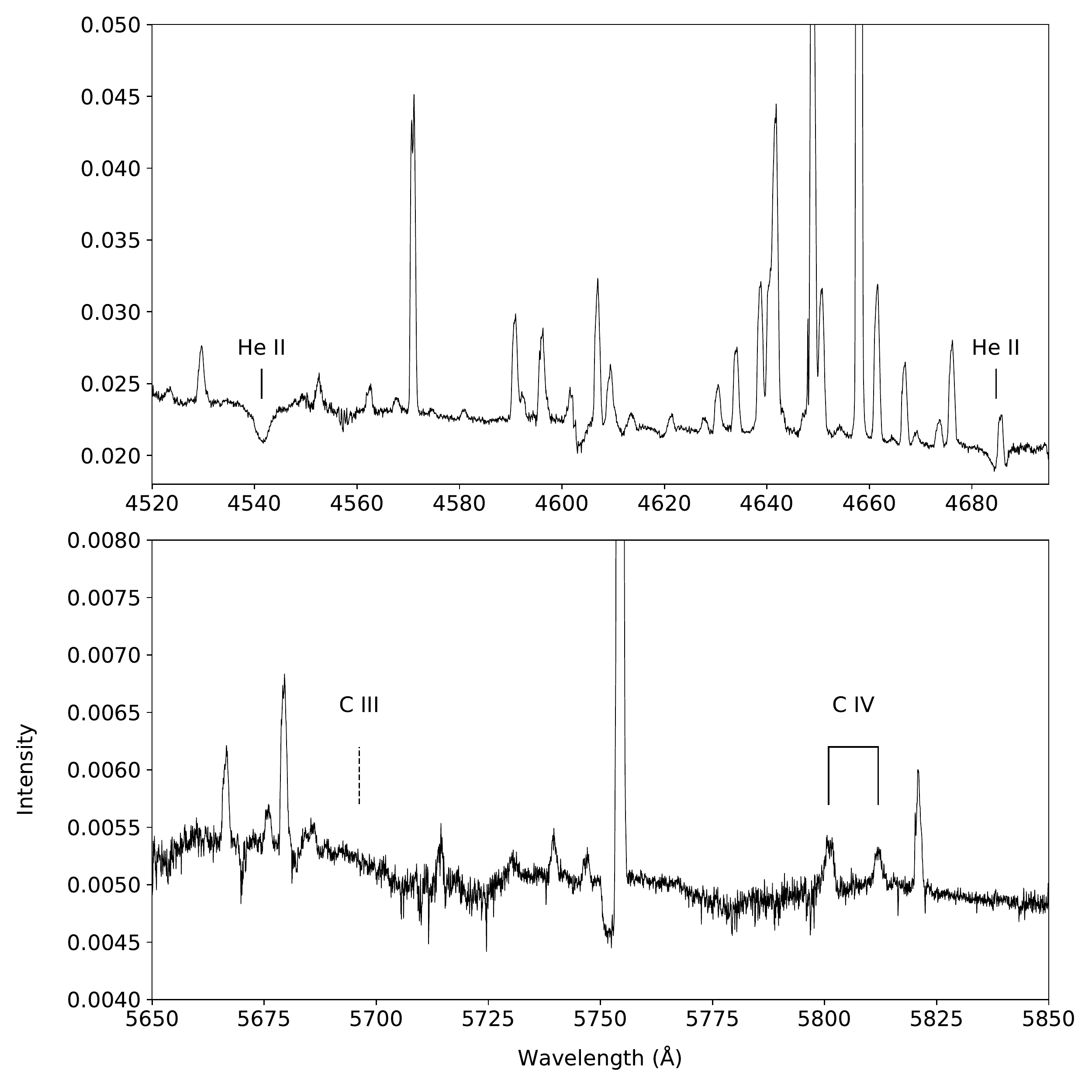}
   \end{center}
   \caption{Portions of the average SALT HRS spectrum corrected for orbital motion. Key features labelled include He~II $\lambda$4541 and $\lambda$4686 (both in absorption) and very weak C~IV $\lambda$5801, 5812. The dashed line shows the expected position of C~III $\lambda$5696 at the heliocentric RV of the nebula which is absent. The spectrum rules out all Wolf-Rayet or Of-type classifications for the central star. Nebular lines appear broadened here due to the orbital motion correction and the broad variations in the red continuum are due to imperfect order merging.}
   \label{fig:spec}
\end{figure}

Figure \ref{fig:spec} shows portions of an average SALT HRS spectrum that was created by correcting each spectrum for the orbital motion (Tab. \ref{tab:log}) before combining all spectra. The absence of C~III $\lambda$5696, together with He~II $\lambda$4541 and $\lambda$4686 both in absorption and very weak C~IV $\lambda$5801, 5812, all rule out classification as either a Wolf-Rayet (e.g. Crowther et al. 1998; Acker \& Neiner 2003) or Of (M\'endez et al. 1990) type central star. This is consistent with the Herald \& Bianchi (2004) non-Wolf-Rayet interpretation of the UV spectrum. Model atmosphere analysis of the UV spectrum by Herald \& Bianchi (2004) determined a surface gravity of $\log g=5.1$ cm s$^{-2}$ and $T_\mathrm{eff}=60^{+10}_{-5}$ kK, consistent with the young kinematic age of the nebula (Sowicka et al. 2017). Herald \& Bianchi (2004) were not able to determine whether the central star was H-rich or H-deficient. The former would correspond to an O(H) classification (M\'endez 1991), while the latter might indicate a very rare O(He) classification (e.g. Reindl et al. 2014 and ref. therein). Further investigation into the atmospheric composition of the primary would require higher quality, high-resolution UV spectra than currently available. Lastly, we note that the central star does still have a substantial wind ($v_\infty=2300\pm200$ km s$^{-1}$, Herald \& Bianchi 2004) and this may help explain the relatively large residuals of 4.1 km s$^{-1}$ observed in the RV measurements (Sect. \ref{sec:orbit}).

\section{CONCLUSIONS}
\label{sec:conclusion}
We have obtained and analysed 16 SALT HRS observations of the central star of IC~4776 ($R=\lambda/\Delta \lambda=43000$). Periodogram analysis of SALT HRS RV measurements of the He~II $\lambda$4541 absorption line result in a 3.11 d orbital period, significant at the 5$\sigma$ level, proving the post-CE binary nature of the central star. Assuming a primary mass of 0.6 $M_\odot$ and an orbital inclination matching the nebula orientation of $48\pm4$ deg (Sowicka et al. 2017), the mass function results in a companion mass of $0.22\pm0.03$ $M_\odot$, most likely corresponding to an M4V star. The mass function also rules out a WD companion more massive than the primary. The orbital period is consistent with the expected bias-corrected orbital period distribution of WDMS binaries (Nebot G\'omez-Mor\'an et al. 2011). However, long orbital period binary central stars of PNe (e.g. Miszalski et al. 2018b; Brown et al. 2018) remain a promising avenue to constrain CE population synthesis models that incorporate recombination energy in the CE ejection (Zorotovic et al. 2014a). The conclusions of Wesson et al. (2018) regarding low ADF values observed in longer orbital periods are not significantly changed, however there remain strong selection effects influencing any inferred correlations. We rule out all Wolf-Rayet and Of classifications for the primary after building a deep stack of SALT HRS spectra corrected for orbital motion. Higher quality UV spectra are required to further investigate the atmospheric composition of the primary.

\section{Acknowledgements}
BM acknowledges support from the National Research Foundation (NRF) of South Africa and thanks the Institute of Astronomy at KU Leuven for their hospitality. We thank the anonymous referee for a helpful report whose comments helped improve this paper. This paper is based on spectroscopic observations made with the Southern African Large Telescope (SALT) under programme 2017-1-MLT-010 (PI: B. Miszalski). We are grateful to our SALT colleagues for maintaining the telescope facilities and conducting the observations. We thank A. Y. Kniazev for making available his HRS pipeline data products. HVW and RM acknowledge support from the Belgian Science Policy Office under contract BR/143/A2/STARLAB. HVW acknowledges additional support from the Research Council of K.U. Leuven under contract C14/17/082. Polish participation in SALT is funded by grant No. MNiSW DIR/WK/2016/07. This research has been partly founded by the National Science Centre, Poland, through grant OPUS 2017/27/B/ST9/01940 to JM. IRAF is distributed by the National Optical Astronomy Observatory, which is operated by the Association of Universities for Research in Astronomy (AURA) under a cooperative agreement with the National Science Foundation.

\end{document}